\RequirePackage{snapshot}       
\documentclass[11pt,a4paper]{article}
\usepackage[utf8]{inputenc}
\usepackage{cite}
\usepackage{amsmath,amsfonts,amssymb}
\usepackage{graphicx}
\usepackage{psfrag}
\usepackage{multicol}
\usepackage{dsfont,mathabx}
\usepackage{fullpage}

\usepackage{bbm}

\usepackage{lmodern}
\usepackage{empheq}
\usepackage{authblk}

\newtheorem{definition}{Definition}

\newtheorem{theorem}[definition]{Theorem}

\newcommand{\poly}{ {\mathrm{poly}}}

\title{\textbf{Probabilistic Self-Stabilization}  \\
 {\small (Communication) }}
  
\date{February 19, 2015}

\author[1]{L. Becchetti}
\author[2]{A. Clementi}
\author[1]{E. Natale}
\author[2]{F. Pasquale}

\affil[1]{\emph{Sapienza} Universit\`a di Roma, {\tt natale@di.uniroma1.it}}
 
\affil[2]{Universit\`a \emph{Tor Vergata} di Roma, {\tt clementi@mat.uniroma2.it}, {\tt pasquale@mat.uniroma2.it}}

\begin{document}
\thispagestyle{empty}
\maketitle
\begin{abstract}
By using concrete scenarios, we
 present and discuss a new concept of \emph{probabilistic Self-Stabilization} in \emph{Distributed Systems}.

\end{abstract}

\section{Introduction}

A \emph{distributed system} is a network of \emph{agents} that 
coordinate their actions by exchanging messages~\cite{Peleg00}. In 
order to be effective, distributed systems have to be \emph
{self-stabilizing}~\cite{Dijkstra74}: A system is self-stabilizing 
if, starting from an arbitrary state, it will quickly reach a \emph
{legitimate} state and once in a legitimate state, it will only 
take on legitimate states thereafter. Self-stabilization has important consequences:
for example it allows  (fast) recovery when the system is prone to \emph{transient faults} 
that might take into non-legitimate states \cite{D74,L85,D00}. Self-stabilization is a 
mature subject in the area of Distributed Computing~\cite{Dolev00} 
and self-stabilizing algorithms for classical computational tasks 
are nowadays well-understood.

The original concept of self-stabilization is too restrictive to 
properly describe some modern systems, e.g., P2P and social networks, 
which  are \emph{dynamic}. Indeed, several relaxations have  
been proposed so far: \emph{probabilistic self-stabilization}~\cite
{IJ90}, where randomized strategies for self-stabilization are 
allowed; \emph{pseudo self-stabilization}~\cite{bgm93}, where the 
system is allowed to deviate from legitimate states for a finite 
amount of time; \emph{$k$-self-stabilization}~\cite{bgk98}, where 
restrictions on the initial state are imposed (namely, all allowed 
initial states are those from which a legitimate state of the system 
can be reached by changing the state of at most $k$ agents); \emph
{weak self-stabilization}~\cite{Gouda01}, that only requires the 
\emph{existence} of an execution that eventually converges to a 
legitimate state.  However,

\begin{center}
\begin{minipage}{0.7\textwidth}
all the above relaxations fail to capture the notion of a 
system that is self-stabilizing only \emph{with high probability} 
and that is required  to remain in legitimate states only over a \emph
{sufficiently long time interval}.
\end{minipage}
\end{center}

The main goal of this work is to discuss a new \emph
{probabilistic} notion of self-stabilization that is general 
enough to apply to a wide class of complex distributed systems and 
suitable to derive algorithmic principles that induce 
effective and useful self-stabilizing behavior in such systems. 
In general, we say a system is self-stabilizing, according to this revised
notion, if

\begin{description}
\item[a.] From any state it will (quickly) converge to a legitimate state \emph{with high probability (w.h.p.)}, and
\item[b.] Once in a legitimate state, w.h.p. it will only take on legitimate states 
over a sufficiently long time span (for instance, over an arbitrarily-large polyinomial number of steps).
\end{description}

In the next section, we illustrate this new concept in reference to a 
specific and important scenario, in which probabilistic 
self-stabilization turned out to be an effective tool of analysis on 
one hand and naturally led to challenging open questions on the other.

\section{Self-Stabilizing Repeated Balls-into-Bins}

We consider  the following   \emph{repeated  balls-into-bins} 
process over an undirected graph $G(V,E)$ with  $|V|=n$ nodes. 
Initially,  $n$ balls are assigned to the $n$ nodes  in an arbitrary way.  
Then, in every round, one   ball is  
chosen from each non-empty node according to  some strategy 
(random, FIFO, etc) and  re-assigned  to one of the node's neighbours uniformly 
at random. Thus, at every time $t$ each node has a (possibly empty) queue of balls waiting 
to be forwarded, while every ball performs a sort of \emph{delayed} random walk over the  
graph, the delay of each random walk depending on the sizes of the 
queues it encounters along its path. It thus follows that 
these random walks are correlated. The main issue here is to 
investigate the impact of such correlation on the maximum load.

Inspired by previous concepts of (load) stability \cite{AKU05,BFG03}, 
we study     the  \emph{maximum load} $M^{(t)}$, i.e.,  the maximum 
queue size at  round $t$ and we are 
interested in the largest $M^{(t)}$ achieved by the process over a 
period of (any) \emph{polynomial} length. 
In the rest of this section 
we assume $G$ is the complete graph.

\paragraph{Applying the notion of probabilistic self-stabilization.}
We next discuss an approach that relies on the notion of probabilistic 
self-stabilization to bound the maximum load, also resulting in tighter 
bounds to the parallel cover time on 
the complete graph. In this approach, the state of the process at 
any time $t$ is completely specified by its {\em configuration}, 
specifying the queue size of each node at time $t$.\footnote{Note 
that, at least to characterize the maximum load, we can assume that balls are 
indistinguishable.} Our notion
of probabilistic self-stabilization discussed above lends itself to a natural
characterization of this process, for which it specializes as follows:

\begin{definition}{\bf (Self-Stabilizing Repeated Balls into 
Bins.)}\label{def:ssrb}
\begin{itemize}
\item
A configuration is  \emph{legitimate}  if its maximum load is 
$O(\log n)$ and a process is \emph{stable} if, starting from 
any legitimate configuration, it only takes on  legitimate 
configurations over a period of $\poly(n)$ length, w.h.p.
\item A   process  is  \emph{self-stabilizing}
if it is stable and if, moreover, starting from \emph{any} 
configuration, it reaches a legitimate configuration, w.h.p.
\item 
The  \emph{convergence time} of a self-stabilizing process is the 
maximum number of rounds required to reach a legitimate configuration starting  
from any configuration.
\end{itemize}
\end{definition}

\noindent
It is important to observe that, unlike    previous concepts of self-stabilization, here there is
always a small chance that 
the system   leaves   legitimate states even if no ``external'' events (e.g. faults)  do happen.
This natural notion of (probabilistic) self-stabilization was also
inspired  by the one proposed  in  \cite{IJ90} for other distributed 
processes. 
On the other hand, stability   impacts  other  important aspects of 
this process. For instance, if the process is stable, we can prove 
good upper bounds on  the   \emph{progress} of a ball, namely, the 
number of   rounds in which the ball  is selected  from its 
current  queue and forwarded, over  a sequence  of $t\geqslant 1$ rounds. 
In turn, this provides a useful tool to bound the \emph{parallel} cover 
time, i.e.,   the time required for  every   ball to visit \emph
{all} nodes (further details about this issue are  given below along   this section).

\paragraph{Repeated-balls-into-bins: past work.} 
To the best of our knowledge, the repeated balls-into-bins process 
was first studied    in \cite{BCE10} where it is used 
as a crucial sub-procedure to 
optimize  the message complexity of a  gossip algorithm over the 
complete graph. The  previous  analysis in \cite{BCE10,EK15}   
(only) holds over very-short (i.e. logarithmic)  periods.
On the other hand,
analysis in \cite{BCN14} considers periods of arbitrary length, but 
it (only) yields a bound on the maximum load that rapidly 
increases with time: after  $t$ rounds, the maximum load is 
$O\!\left(\sqrt t \right)$  w.h.p.
By adopting the FIFO strategy at every bin queue,     the latter 
result easily implies that   the  progress of any ball over $t$ 
consecutive rounds is $\Omega(\sqrt t)$ w.h.p.  
Moreover, it is well known that the cover time for the single-ball 
process is w.h.p. $\Theta(n \log n)$ (it is in fact equivalent to 
the \emph{coupon's collector} process \cite{MU05}).  These  two 
facts easily imply an upper bound  $O\!\left( n^2 \log^2 n 
\right)$ for the parallel cover time of the repeated balls-into-bins 
process on the complete graph.

In this respect, previous analyses of the maximum load   in \cite
{BCN14,BCE10,EK15} are far from tight, since they   rely on  
rough approximations of the process via  other, much simpler 
Markov chains: for instance, in \cite{BCN14}, the authors consider 
the   process - which obviously dominates the original one - where, 
at the beginning of every round,  a new ball  is added to  every  empty bin. 
Clearly, this approach  does not exploit a key global invariant 
(the fixed number $n$ of balls) of the original process.
Previous results are thus not helpful to establish  whether  this 
process is stable (or, even more, self-stabilizing).

In \cite{BCNPP05},  our group proposed a new, tight 
analysis of the repeated balls-into-bins process that significantly 
departs from previous ones, showing that 
the system is     self-stabilizing in the sense of Definition \ref{def:ssrb}.   
These results are summarized in the following
 
\begin{theorem} [\cite{BCNPP05}]\label{thm::main} Let $c$ be an 
arbitrarily-large constant, and let the process start from any 
legitimate configuration. The maximum load $M^{(t)}$      is  
$O(\log n)$ for all  $t = O(n^c)$, w.h.p. 
Moreover, starting from any configuration,  the system 
reaches a legitimate configuration within $O(n)$ rounds, 
w.h.p. 
\end{theorem}
 
\noindent The  above  result strongly improves over the best 
previous bounds  \cite{BCN14,BCE10,EK15} and it is almost tight 
(since we know that maximum load is $\Omega(\log n /\log\log n)$ at 
least during the first rounds \cite{RS98}). Moreover, under the FIFO 
forwarding strategy, the progress 
of any ball over a sequence of $t = \poly(n)$ 
consecutive rounds is $\Omega(t/\log n)$ w.h.p. Consequently, 
the parallel  cover time  on the complete graph   is  $O\!(n \log^2n)$ 
w.h.p., which is  only a $\log n$ factor away from the lower   bound  
following  from   the single-ball process.

\paragraph{Wrapping up: balls-into-bins, distributed computing and 
self-stabilization.}
As mentioned above, besides being interesting \emph{per-se}, balls-into-bins processes  are 
used to model and analyze   several important randomized protocols 
in parallel and distributed computing \cite{ABKU99,BCSV06,V03}. In 
particular,  the process we study  models a  natural 
randomized solution to the problem of \emph{(parallel) resource 
(or task) assignment}   in distributed systems  (this problem is 
also known as  \emph{traversal}) \cite{S06,L86}. For a more detailed discussion
of this potential application, the reader may refer to \cite{BCNPP05}. 

 Theorem  
\ref{thm::main} also allows to study the adversarial model in 
which, on some \emph{faulty} rounds, an adversary can re-assign  
balls to the bins in an arbitrary way. The self-stabilization property and 
the  linear convergence time shown in Theorem~\ref{thm::main}  in fact ensure 
that the  $O(n \log^2n)$ upper bound on the cover time still   holds, 
provided faulty rounds occur with a frequency no higher than 
$c n$, for a sufficiently small constant $c$.

\subsection{Open Questions}
Below, we briefly discuss two challenging extensions to the
basic problem discussed above, that appear natural in the probabilistic 
self-stabilization setting we defined.

\paragraph{More  balls.}  
Consider the Repeated Balls-into-Bins process with $m > n$ balls and $n$ 
bins and  define a state to be \emph{legitimate} if the maximum load 
is $O(m/n \cdot \log n)$ (or, even better, $O(m/n + \log n)$). 
Then,   we can pose the  same  questions: Is  the system 
self-stabilizing? Can we get a good upper bound on the convergence 
time? Experimental results performed  on systems with parameters $m 
\sim n \log n$ (and increasing values of $n$ up to 1 million) seem 
to suggest that this might be the case. Also and quite 
surprsingly, starting from any initial state, the system seems to quickly reach 
and remain confined to configurations that result in a constant fraction 
of empty bins. On the one hand,  if proved, this property would drastically distinguish the 
behaviour of this process from the memoryless one in which, in each round, 
\emph{all} $m$ balls are randomly assigned to the $n$ 
bins independently of their current positions (i.e., there are no bin 
queues). Indeed, in the latter process, a well-known result implies that, 
w.h.p., all bins will be non-empty over any  polynomial number of 
rounds. On the other hand, convergence  towards a constant 
fraction of empty bins is a key-ingredient for proving 
self-stabilization in  the case $m=n$ (i.e. Theorem \ref
{thm::main}). Thus, we believe that, at least when $m \leqslant  n 
\log n$, the system is still self-stabilizing.
 
\paragraph{General Graphs.} Analyzing the (probabilistic) 
self-stabilization properties of the repeated-balls-into-bins process 
turned out to be extremely challenging in networks other than the 
clique: This ``hardness'' seems to hold even for  restricted classes of graphs. For instance, we 
tried to extend our analysis to the most symmetric, sparse case: the 
ring. Yet, while intuition and simulation results suggest 
that the process is self stabilizing in the sense defined above, we were
not able to get any rigorous analytical results so far.

\section{Further research directions}
The notion of probabilistic self-stabilization applies to other 
fundamental problems in Distributed Computing. One of the most
interesting is the classic problem of reaching (w.h.p.) a 
\emph{stabilizing  consensus} (or even more a stabilizing \emph
{majority} consensus \cite{AAE07,A12,BCN14,D74,DGMSS11}) in a 
distributed system consisting of a finite sets of agents. In a 
basic variant, each of the agents initially supports an opinion 
(say a value  chosen from  a fixed set $K$  of legal values). The 
goal here  is have the system converge to a state in which (w.h.p.) 
all nodes share the same opinion and this was present in the initial 
configuration. Moreover, solutions are required to be fault-tolerant  (i.e. stable) w.r.t. 
some bounded adversary that can   change the   values of a subset of 
the nodes in each round of the consensus process. Important advances in probabilistic 
versions of stabilizing consensus were recently made \cite
{AAE07,BCNPST13-full,BCN14,DGMSS11}. However, available  concepts 
of  stabilizing consensus do not fully reflect our proposed notion   
of probabilistic  \emph{self-stabilization}. Our weaker version of consensus may lead to 
more efficient and more robust protocols that still work in practice for most 
applications.





\bibliographystyle{abbrv}
\bibliography{congestion}

\begin{thebibliography}{10}

\bibitem{AKU05}
A.~Anagnostopoulos, A.~Kirsch, and E.~Upfal.
\newblock Load balancing in arbitrary network topologies with stochastic
  adversarial input.
\newblock {\em SIAM Journal on Computing}, 34(3):616--639, 2005.

\bibitem{AAE07}
D.~Angluin, J.~Aspnes, and D.~Eisenstat.
\newblock A simple population protocol for fast robust approximate majority.
\newblock In {\em Proc. of 21st DISC}, volume 4731 of {\em LNCS}, pages 20--32.
  Springer, 2007.

\bibitem{A12}
J.~Aspnes.
\newblock Faster randomized consensus with an oblivious adversary.
\newblock In {\em PODC'12}. ACM, 2012.

\bibitem{ABKU99}
Y.~Azar, A.~Z. Broder, A.~R. Karlin, and E.~Upfal.
\newblock Balanced allocations.
\newblock {\em SIAM journal on computing}, 29(1):180--200, 1999.

\bibitem{bgk98}
J.~Beauquier, C.~Genolini, and S.~Kutten.
\newblock \emph{k}-stabilization of reactive tasks.
\newblock In {\em Proceedings of the Seventeenth Annual {ACM} Symposium on
  Principles of Distributed Computing, {PODC} '98, Puerto Vallarta, Mexico,
  June 28 - July 2, 1998}, page 318, 1998.

\bibitem{BCNPP05}
L.~Becchetti, A.~Clementi, E.~Natale, F.~Pasquale, and G.~Posta.
\newblock Self-stabilizing repeated balls-into-bins.
\newblock In {\em 27th ACM Symposium on Parallelism in Algorithms and
  Architectures}, 2015.
\newblock To appear.

\bibitem{BCN14}
L.~Becchetti, A.~Clementi, E.~Natale, F.~Pasquale, and R.~Silvestri.
\newblock Plurality consensus in the gossip model.
\newblock In {\em Proceedings of the 25th {ACM-SIAM} Symposium on Discrete
  Algorithms (SODA)}, 2015.

\bibitem{BCNPST13-full}
L.~Becchetti, A.~Clementi, E.~Natale, F.~Pasquale, R.~Silvestri, and
  L.~Trevisan.
\newblock Simple dynamics for plurality consensus.
\newblock {\em arXiv:1310.2858}, 2013.
\newblock Preliminary version in ACM SPAA'14.

\bibitem{BCSV06}
P.~Berenbrink, A.~Czumaj, A.~Steger, and B.~V{\"o}cking.
\newblock Balanced allocations: The heavily loaded case.
\newblock {\em SIAM Journal on Computing}, 35(6):1350--1385, 2006.

\bibitem{BCE10}
P.~Berenbrink, J.~Czyzowicz, R.~Els{\"a}sser, and L.~Gasieniec.
\newblock Efficient information exchange in the random phone-call model.
\newblock In {\em Proceedings of the 37th International Colloquium on Automata,
  Languages, and Programming (ICALP)}, 2010.

\bibitem{BFG03}
P.~Berenbrink, T.~Friedetzky, and L.~A. Goldberg.
\newblock The natural work-stealing algorithm is stable.
\newblock {\em SIAM Journal on Computing}, 32(5):1260--1279, 2003.

\bibitem{bgm93}
J.~E. Burns, M.~G. Gouda, and R.~E. Miller.
\newblock Stabilization and pseudo-stabilization.
\newblock {\em Distributed Computing}, 7(1):35--42, 1993.

\bibitem{Dijkstra74}
E.~W. Dijkstra.
\newblock Self-stabilizing systems in spite of distributed control.
\newblock {\em Commun. {ACM}}, 17(11):643--644, 1974.

\bibitem{D74}
E.~W. Dijkstra.
\newblock Self-stabilizing systems in spite of distributed control.
\newblock {\em Communications of the ACM}, 17(11):643--644, 1974.

\bibitem{DGMSS11}
B.~Doerr, L.~A. Goldberg, L.~Minder, T.~Sauerwald, and C.~Scheideler.
\newblock Stabilizing consensus with the power of two choices.
\newblock In {\em Proc. of 23rd SPAA}, pages 149--158. ACM, 2011.

\bibitem{D00}
S.~Dolev.
\newblock {\em Self-stabilization}.
\newblock MIT press, 2000.

\bibitem{Dolev00}
S.~Dolev.
\newblock {\em Self-Stabilization}.
\newblock The MIT Press, 2000.

\bibitem{EK15}
R.~Els{\"a}sser and D.~Kaaser.
\newblock On the influence of graph density on randomized gossiping.
\newblock {\em Proceedings of the 29th IEEE International Parallel $\&$
  Distributed Processing Symposium (IPDPS)}, 2015.

\bibitem{Gouda01}
M.~G. Gouda.
\newblock The theory of weak stabilization.
\newblock In {\em Self-Stabilizing Systems, 5th International Workshop, {WSS}
  2001, Lisbon, Portugal, October 1-2, 2001, Proceedings}, pages 114--123,
  2001.

\bibitem{IJ90}
A.~Israeli and M.~Jalfon.
\newblock Token management schemes and random walks yield self-stabilizing
  mutual exclusion.
\newblock In {\em Proceedings of the 9th annual ACM Symposium on Principles of
  Distributed Computing (PODC)}, 1990.

\bibitem{L85}
L.~Lamport.
\newblock Solved problems, unsolved problems and non-problems in concurrency.
\newblock {\em ACM SIGOPS Operating Systems Review}, 19(4):34--44, 1985.

\bibitem{L86}
N.~A. Lynch.
\newblock {\em Distributed algorithms}.
\newblock Morgan Kaufmann, 1996.

\bibitem{MU05}
M.~Mitzenmacher and E.~Upfal.
\newblock {\em Probability and computing: Randomized algorithms and
  probabilistic analysis}.
\newblock Cambridge University Press, 2005.

\bibitem{Peleg00}
D.~Peleg.
\newblock {\em Distributed Computing: A Locality-Sensitive Approach}.
\newblock SIAM, 2000.

\bibitem{RS98}
M.~Raab and A.~Steger.
\newblock “balls into bins”—a simple and tight analysis.
\newblock In {\em Proceedings of the 2nd International Workshop on
  Randomization and Approximation Techniques in Computer Science (RANDOM)},
  1998.

\bibitem{S06}
N.~Santoro.
\newblock {\em Design and analysis of distributed algorithms}.
\newblock John Wiley \& Sons, 2006.

\bibitem{V03}
B.~V{\"o}cking.
\newblock How asymmetry helps load balancing.
\newblock {\em Journal of the ACM}, 50(4):568--589, 2003.

\end{thebibliography}


\end{document}